\documentclass[twocolumn,prl,showpacs,superscriptaddress]{revtex4}

\usepackage{graphicx}
\usepackage{dcolumn}
\usepackage{bm}

\begin{document}

\title{Particle-Hole Symmetry and the Fractional Quantum Hall States at $5/2$ Filling Factor}
\author{Jian Yang}
\email{jyangmay1@yahoo.com}
\altaffiliation{Permanent address: 5431 Chesapeake Place, Sugar Land, TX 77479, USA}

\begin{abstract}

We propose a derivative operator formed as a function of derivatives of electron coodinates defined by
$D_m = {Pf} ( \frac{1}{ (\frac{\partial}{{\partial}z_i}-\frac{\partial}{{\partial}z_j})^m } )  {\prod}_{i<j}^N(\frac{\partial}{{\partial}z_i}-\frac{\partial}{{\partial}z_j} ) ^m$, with $z_j$ being the complex coordinate of the $j_{th}$ electron, $N$ the total number of electrons, $m$ a positive integer, and $Pf[A]$ is the Pfaffian of an antisymmetric matrix $A$. When applied to the Laughlin wave function 
${\Phi}_{m_L}  = {\prod}_{i<j}^N (z_i-z_j)^{m_L}$, 
a new wave function ${\Psi}_{m,m_L} =  D_m {\Phi}_{m_L}$ in the lowest Landau level at filling factor $\nu = 1/(m_L-m)$ is generated. In spherical geometry,   the relationship between the total magnetic flux number $N_{\phi}$ and $N$ is $N_{\phi}= (m_L-m)N+(2m-m_L)$. Two wave functions 
${\Psi}_{3,5}$ and ${\Psi}_{1,3}$ with special sets of values $(m,m_L) = (3,5)$ and $(m,m_L) = (1,3)$, are of particular interest as they both correspond to a half-filled Landau level and are relevant to the $5/2$ quantum Hall effect. The first wave function ${\Psi}_{3,5}$
has the $N_{\phi}$ and $N$ relationship $N_{\phi} = 2N+1$, and the second wave function ${\Psi}_{1,3}$ has $N_{\phi} = 2N-1$. 
For systems of 4, 6, and 8 electrons in spherical geometry, it is shown that the 
first wave function ${\Psi}_{3,5}$ has nearly unity overlap with the particle-hole conjugate of the Moore-Read Pfaffian wave function, therefore together
with the Moore-Read Pfaffian state forms a particle-hole conjugate pair. The second wave function ${\Psi}_{1,3}$ has essentially perfect particle-hole symmetry itself, with a positive parity when the number of electron pairs $N/2$ is an even integer and and a negative parity when $N/2$ is an odd integer. An equivalent form suggests the first wave function ${\Psi}_{3,5}$ forms a f-wave pairing of composite fermions, and the second wave function ${\Psi}_{1,3}$ forms a p-wave pairing. The corresponding Non-Abelian statistics quasiparticle wave functions are also proposed. 
\end{abstract}
 \maketitle

It has been thirty years since the observation of the fractional quantum Hall effect at
$\nu=5/2$\cite{Willett}, which deviates from the odd denominator
filling factor rule. The Moore-Read Pfaffian state (termed the Pfaffian state in the literature)\cite{MR} is considered to
be the leading candidate for the ground state, and has been studied extensively in particular
through finite size numerical calculations\cite{Morf}\cite{Rezayi}\cite{Peterson}. It is shown
by adjusting the pseudopotentials \cite{Morf} or fine tuning the finite thickness of the layer that confined the two
dimensional electrons\cite{Peterson}, the ground state of the
resulting Columb Hamiltonian at the second Landau level can achieve nearly unity overlap with the wave function of the Pfaffian state. 

It is known that the Pfaffian state is the exact ground state of a three-body interaction Hamiltonian\cite{Greiter}, 
and therefore does not have particle-hole (PH) symmetry. On the other hand, the two-body interaction Hamitonian 
projected into a single Landau level is invariant by an antiunitary PH transformation, which requires its ground state to be PH symmetric. 
This apparent paradox can be resolved in two ways. One is to invoke a PH symmetry breaking mechanism such as 
Landau level mixing, which lifts the degeneracy between the ground state described by the Pfaffian state and 
the other degenerate ground state described by the distinct PH conjugate of the Pfaffian state, termed anti-Pfaffian state in the literature,
such that one of them becomes the ground state\cite{Levin}\cite{Lee}. The other resolution is the PH symmetry is preserved, and neither the Pfaffian state nor the anti-Pfaffian state provides a good description of the ground state. Rezayi and Haldane \cite{Rezayi} in torus geometry provided evidence that the numerical ground state that corrsponds to the $5/2$ quantum Hall effect does have PH symmetry, and is identified as a symmetrized superposition of the Pfaffian state and the anti-Pfaffian state (see also \cite{Wang}). A possibility of spontanouse PH symmetry breaking is also studied and ruled out for the Coulomb interaction even when the finite-thickness effects are included\cite{DasSarma}. Recently, a Dirac composite fermion effective field theory has been proposed to describe the low energy physics where the PH symmetry is explicitly realized\cite{Son}. 

Although much progress has been made, there are two important questions remain unanswered. In the case of the PH symmetry breaking, if we believe
the Pfaffin wave function provides a good description for one of the two degenate states that are PH conjugated with each other, what would the wave function look like for the other state? In the case of the PH symmetry is preservered, can we construct a wave function that is PH symmetric in the first place and has a
large overlap with the exact ground state as well as the correct parity. We wish to make contributions to the answers to both questions in this paper.

We begin by defining a derivative operator $D_m$ in the planar geometry as follows:
\begin{equation}
\label{DOperator} D_m = {Pf} (\frac{1}{ (\frac{\partial}{{\partial}z_i}-\frac{\partial}{{\partial}z_j})^m }) \prod\limits_{i<j}^N(\frac{\partial}{{\partial}z_i}-\frac{\partial}{{\partial}z_j} ) ^m
\end{equation}
where $z_j = x_j+iy_j$ is the complex coordinate of the $j_{th}$ electron, $N$ is the total number of electrons, $m$ is a positive integer, and $Pf[A]$ is the Pfaffian of an antisymmetric matrix $A$. When applied to the Laughlin wave function the following new wave function ${\Psi}_{m,m_L}$ is generated:
\begin{equation}
\label{DWaveFunction} {\Psi}_{m,m_L} =D_m  {\Phi}_{m_L} 
\end{equation}
with the Laughlin wave function ${\Phi}_{m_L}$ defined by\cite{Laughlin}
\begin{equation}
\label{Laughlin} {\Phi}_{m_L}  = \prod\limits_{i<j}^N (z_i-z_j)^{m_L} \exp(-\sum\limits_j \frac{|z_j|^2}{4l_B^2})
\end{equation}
where $m_L > m$ being an odd integer for fermions and an even integer for bosons, and $l_B$ is the magnetic length. It is noted that the product of the Pfaffian and the Jastrow function of the derivatives in Eq.(\ref{DOperator}) should be carried out first before applying to the Laughlin wave function, this way the derivatives appeared in the denominator in the Pfaffian will be cancelled out. It should also be noted that the $D_m$ does not apply to the exponent factor of the Laughlin wave function $\exp(-{\sum}_j \frac{|z_j|^2}{4l_B^2})$. It is clear the wave function ${\Psi}_{m,m_L}$ has the same symmetry as the Laughlin
wave function, which is antisymmetric with respect to the coordinates when $m_L$ is an odd interger, and symmetric when $m_L$ is an even interger. 

In Haldane's spherical geometry\cite{Haldane}, the $D_m$ operator can be written as:
\begin{equation}
\label{DOperator_Sphere} D_m = {Pf}(\frac{1}{ ( \frac{\partial}{{\partial}u_i}\frac{\partial}{{\partial}v_j}-\frac{\partial}{{\partial}u_j}\frac{\partial}{{\partial}v_i})^m })  \prod\limits_{i<j}^N(\frac{\partial}{{\partial}u_i}\frac{\partial}{{\partial}v_j}-\frac{\partial}{{\partial}u_j}\frac{\partial}{{\partial}v_i} ) ^{m}
\end{equation}
and the Laughlin wave function can be written as
\begin{equation}
\label{Laughlin_Sphere} {\Phi}_{m_L} =   \prod\limits_{i<j}^N (u_iv_j-u_jv_i)^{m_L} 
\end{equation}
where $(u, v)$ are the spinor variables describing electron coordinates. Since the total flux number $N_{\phi}$ corresponding to the Laughlin wave function is $N_{\phi} = m_L(N-1)$, and the derivative operator $D_m$ decreases the flux number by $m(N-2)$, the relationship between the flux number $N_{\phi}$ and the number of electrons $N$ is given by the following equation: 
\begin{equation}
\label{FluxNRelation}N_{\phi} = (m_L-m)N+(2m-m_L)
\end{equation}
for the wave function ${\Psi}_{m,m_L}$. In the thermodynamic limit, this corresponds to filling factor $\nu = 1/(m_L-m)$.  

In the following, we will focus on the two wave functions ${\Psi}_{3,5}$ and ${\Psi}_{1,3}$ described by Eq.(\ref{DWaveFunction}) with special sets of values $(m,m_L) = (3,5)$ and $(m,m_L) = (1,3)$. Both wave functions correspond to filling factor $\nu = 1/2$. 

According to Eq.(\ref{FluxNRelation}), the wave function ${\Psi}_{3,5}$ has the $N_{\phi}$ and $N$ relationship $N_{\phi} = 2N+1$. Since the number of
electrons $N$ is related to the numeber of holes $N_h$ of the PH conjugate state by $N+N_h = N_{\phi}+1$, the relationship between the flux number and 
the number of holes of the PH conjugate state of ${\Psi}_{3,5}$, represented by ${\Psi}_{3,5}^{QH}$ hereafter, is $N_{\phi} = 2N_h-3$. This is exactly the same
as the $N_{\phi}$ and $N$ relationship for the Pfaffian state, which is $N_{\phi} = 2N-3$. In other words, for the 
same flux number $N_{\phi}$, the number of holes of the ${\Psi}_{3,5}^{QH}$ state is the same as the number of electrons of the Pfaffian state. 
By the same token, the number of electrons of the ${\Psi}_{3,5}$ state is the same as the number of holes of the ant-Pfaffian state. Encourged by this fact, 
we have explicitly expanded the wave function ${\Psi}_{3,5}$ and represented it in terms of the many-body basis functions formed of the Slater determinant of $N$ single particle wave functions in the lowest Landau level for the small number of electron systems of $N = 4$ and $N = 6$. This representation, as will be illustrated in detail later, makes it very easy to apply the PH transformation to obtain the explicit form of ${\Psi}_{3,5}^{QH}$, and to calculate the overlap between ${\Psi}_{3,5}^{QH}$ and the Pfaffian 
state. The overlap with the Pfaffian state is $0.9997$ for $N = 6$, and $0.9962$ for $N=8$. 

\begin{figure}[tbhp]
\includegraphics[width=\columnwidth]{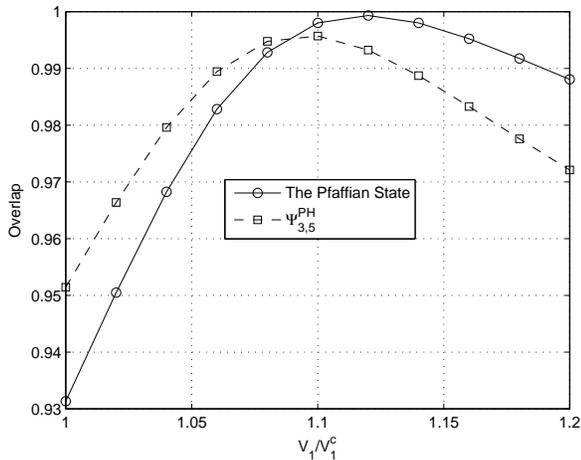}
\caption{\label{fig:Overlap} For $N=8$ and $N_{\phi} = 13$. Overlap of the exact ground state with the Pfaffian state (solid line) and the 
${\Psi}_{3,5}^{PH}$ state (dashed line) as the function of the pseudopotential $V_1$
normalized by its Coulomb value $V_1^c$ in the second Landau level.}
\end{figure}

This nearly unity overlap provides
an evidence that the ${\Psi}_{3,5}^{QH}$ state and the Pfaffian state, or the ${\Psi}_{3,5}$ state and the anti-Pfaffian state, form PH conjugated pair states with each other. Between the Pfaffian state and the ${\Psi}_{3,5}$ state, if one provides a good description for the ground state of the two body interaction, so does the PH conjugate state of the other. In Fig.~\ref{fig:Overlap}, we show the overlap of the exact ground state of the Coulomb interaction with the Pfaffian state and the ${\Psi}_{3,5}^{PH}$ state respectively for ratios of $V_1/V_1^c$ ranging from $1$ to $1.2$, where $V_1^c$ is the Coulomb value of $V_1$ in 
the second Landau level. Overall, both states provide rather good description of the exact ground state. When $V_1$ has the value for Coulomb interaction $V_1^c$, ${\Psi}_{3,5}^{PH}$ has larger overlap than the Pfaffian state. As $V_1$ increases, the overlap for both wave functions increases until the overlap with ${\Psi}_{3,5}^{PH}$ reaches its maximum value around $V_1/V_1^c = 1.1$ and the overlap
with the Pfaffian state reaches its maximum value around $V_1/V_1^c = 1.12$.

Now we turn our attention to the wave function ${\Psi}_{1,3}$. According to Eq.(\ref{FluxNRelation}), the wave function ${\Psi}_{1,3}$ has the $N_{\phi}$ and $N$ relationship $N_{\phi} = 2N-1$. Since the number of
electrons $N$ is related to the numeber of holes $N_h$ of the PH conjugated state by $N+N_h = N_{\phi}+1$, the number of electrons of
${\Psi}_{1,3}$ is the same as the number of holes of the PH conjugate state of ${\Psi}_{1,3}$, represented by ${\Psi}_{1,3}^{QH}$ hereafter. Now the question is if the wave function ${\Psi}_{1,3}$ has the PH symmetry? To answer the question, as before we explicitly expand the wave function ${\Psi}_{1,3}$ in terms of the many-body basis functions formed of the Slater determinant of $N$ single particle wave functions in the lowest Landau level for small number of electron systems 
$N = 4$, $N = 6$, and $N = 8$. We then construct the PH conjugate wave function ${\Psi}_{1,3}^{PH}$ by applying the PH transformation, and calculate
the overlap between ${\Psi}_{1,3}$ and ${\Psi}_{1,3}^{PH}$. To illustrate how this is done, we use the system $N = 4$ as an example, 
as this is the simplest system and the wave function ${\Psi}_{1,3}$ is shown to have the exact PH symmetry. For $N=4$ system, the total flux number is $2N-1 = 7$. In the spherical geometry, the single particle wave function in the lowest Landau Level is specified by the quantum number $m$ of the angular momentum $L_z$ that takes $N_{\phi}+1$ values of $-\frac{N_{\phi}}{2},-\frac{N_{\phi}}{2}+1, {\ldots}, \frac{N_{\phi}}{2}$. For the notation convenience, we will use $\frac{N_{\phi}}{2}+m$ instead which takes values of $0, 1, 2, {\ldots}, N_{\phi}$ to specify the single particle wave function in the following discussion. The Hilbert space of dimension $M$ is spanned by $M$ orthogonal many-body basis functions formed of the Slater determinant of $N$ (in this example $N=4$) single particle wave functions $|i> = |\frac{N_{\phi}}{2}+m_1;\frac{N_{\phi}}{2}+m_2;\frac{N_{\phi}}{2}+m_3;\frac{N_{\phi}}{2}+m_4>$, where $i = 1, 2, \ldots, M$ is used to index the $M$ basis functions. Using the basis function notation, the wave function ${\Psi}_{1,3}$ can be written in the form 
\begin{equation}
\label{SecondQuantizedForm}
{\Psi}_{1,3} = \sum\limits_{i=1}^M C_i |i>.
\end{equation}
where $C_i$ is the coefficient for the basis function $|i>$. 
Since the wave function ${\Psi}_{1,3}$ is rotationally invariant, its total angular momentum is zero. This requires ${\sum_i}m_i = 0$ or equivalently ${\sum_i}(\frac{N_{\phi}}{2}+m_i) = N(2N-1)/2 = 14$. As the result, the total number of the basis functions $M = 8$ for the $4$ electron system. In Table~\ref{tab:table1}, we list all the $8$ basis functions and the corresponding coefficients (unnormalized). We also listed the PH conjugate of each of the basis function $|i>^{PH}$. As can be 
seen, we number the many-body basis functions in an order such that $|i>^{PH}$ is related to $|i>$ by the simple relation $|i>^{PH} = |M-i+1>$. 
In other words, $|i>^{PH}$ can be obtained from $|i>$ by reversing the order. Therefore, the wave function 
${\Psi}_{1,3}^{QH} = {\sum}_i C_i |i>^{QH} = {\sum}_i C_i |M-i+1>$. If we change the index varible $M-i+1 \rightarrow i$, we will have ${\Psi}_{1,3}^{PH} = \sum_{i=1} C_{M-i+1} |i>$.
Since the coefficients as shown in the first coulom in Table~\ref{tab:table1} satisfy the equation $C_{M-i+1} = C_i$, we have ${\Psi}_{1,3}^{QH} = {\sum}_i C_i |i> = {\Psi}_{1,3}$. In other words,  ${\Psi}_{1,3}$ has perfect PH symmetry with parity equal to $+1$ for a system of $4$ electrons.

\begin{table}[h]
\centering
\begin{tabular}{|c|c|c|}
\hline
$C_i$&$|i>$ & $|i>^{PH}$  \\
\hline
  -1&$|1>=|0;1;6;7>$ & $|1>^{PH}=|2;3;4;5>$ \\
\hline
  1&$|2>=|0;2;5;7>$ & $|2>^{PH}=|1;3;4;6>$ \\
\hline
  -1&$|3>=|0;3;4;7>$ & $|3>^{PH}=|1;2;5;6>$ \\
\hline
  0&$|4>=|0;3;5;6>$ & $|4>^{PH}=|1;2;4;7>$ \\
\hline
  0&$|5>=|1;2;4;7>$ & $|5>^{PH}=|0;3;5;6>$ \\
\hline
  -1&$|6>=|1;2;5;6>$ & $|6>^{PH}=|0;3;4;7>$ \\
\hline
  1&$|7>=|1;3;4;6>$ & $|7>^{PH}=|0;2;5;7>$ \\
\hline
  -1&$|8>=|2;3;4;5>$ & $|8>^{PH}=|0;1;6;7>$ \\
\hline
\end{tabular}
\caption{\label{tab:table1}Coefficients and the basis functions (and their PH conjugates) as defined in Eq.(\ref{SecondQuantizedForm})
for $N = 4$ system.}
\end{table}

\begin{figure}[tbhp]
\label{fig:Coefficients_6e}
\includegraphics[width=\columnwidth]{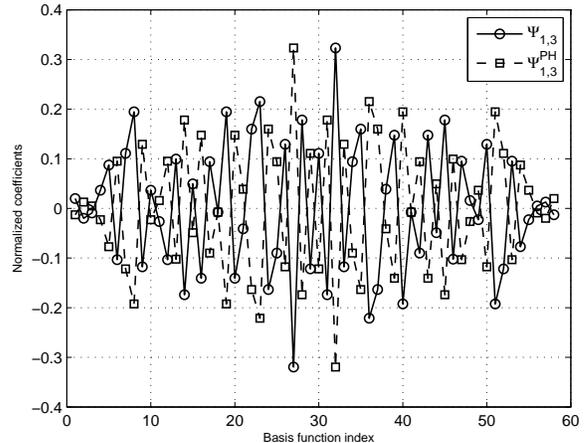}
\caption{\label{fig:Coefficients_6e} The coefficients (cicles) of the wave function ${\Psi}_{1,3}$ and the coefficients (squares) 
of ${\Psi}_{1,3}^{PH}$ for $N=6$ and $N_{\phi} = 11$. The solid line and the dashed line are used to guide the eye.}
\end{figure}

For $N=6$, there are total number of $58$ basis functions. Again as described in Eq.(\ref{SecondQuantizedForm}) we expand the wave function ${\Psi}_{1,3}$ in terms of the basis functions $|i>$ where $i=1,2, {\ldots}, 58$, and calculated the coefficients $C_i$. The normalized coefficients $C_i$ are plotted as cicles  with the solid line used
to guide the eye in Fig.~\ref{fig:Coefficients_6e}. The coefficients of the ${\Psi}_{1,3}^{QH}$ are also plotted as squares with the dashed line used to guide the eye in Fig.~\ref{fig:Coefficients_6e}. It can be seen from the Figure, at each basis function index, ${\Psi}_{1,3}$ and ${\Psi}_{1,3}^{QH}$ have the coefficients that are essentially the same in magitude but in opposite sign, indicating that ${\Psi}_{1,3}$ has the PH symmtry and the parity is $-1$. In
fact, the overlap between ${\Psi}_{1,3}$ and ${\Psi}_{1,3}^{QH}$ is $0.9991$ for $N=6$. The same calculation is also carried out for $N=8$, 
the overlap between ${\Psi}_{1,3}$ and ${\Psi}_{1,3}^{QH}$ is $0.9798$, and the parity is $+1$. These results provide the strong evidence that the following equation
\begin{equation}
\label{PHSymmetric}
{\Psi}_{1,3} = (-1)^{\frac{N}{2}}{\Psi}_{1,3}^{PH}.
\end{equation} 
 is essentially true.

It should be pointed out that in spherical geometry at $N_{\phi} = 2N-1$ the nature of the ground state in the second Landau level seems to be very much $N$ dependent. While we are able to identify a reasonalbe range of values of $V_1/V_1^c$ such that ${\Psi}_{1,3}$ has nearly unity overlap with the exact incompressible ground state with the correct parity for $N=4$ and $N=6$ systems, we have not been able to find such a range of $V_1/V_1^c$ such that the exact ground state is incompressible with the same parity of ${\Psi}_{1,3}$ for $N=8$. It is therefore interesting to find out if ${\Psi}_{1,3}$ can provide a good description (correct parity and large overlap) of the exact ground state when the finite-thickness effects are taken into account to alter all the pseudopotential components instead of just $V_1$\cite{Peterson}. It is also interesting to find out if ${\Psi}_{1,3}$ can provide a good description (correct parity and large overlap) of the exact ground state using torus geometry\cite{Rezayi}, or if the symmetrized/antisymmetrized superposition of the Pfaffian state and ${\Psi}_{3,5}$ state can provide such a description. 

Before closing, we would like to make a few remarks. First, one may contruct the wave function ${\Psi}_{m,m_L}$ in a different way than in Eq.(\ref{DWaveFunction}) as ${\Psi}_{m,m_L} = {\Phi}_{m_L-m-1} D_m  {\Phi}_{m+1}$
with the understanding that $D_m$ only applies to the functions to its right. For example, ${\Psi}_{1,3}$ can be contructed as ${\Phi}_{1} D_1  {\Phi}_{2}$. In fact, we have found the the wave function ${\Phi}_{1} D_1  {\Phi}_{2}$ has an improved overlap over $D_1  {\Phi}_{3}$ with its PH conjugated state to unity from $0.9991$, while the parity remains negative for $N=6$.

Secondly, it is straightfoward to show that ${\Psi}_{3,5}$ can be rewritten in an equivalent form as
\begin{equation}
\label{EquivalentDerivativePfaffian_Sphere} {\Psi}_{3,5} =  P_{LLL} Pf(\frac{1}{(u_i^{\ast}v_j^{\ast}-u_j^{\ast}v_i^{\ast})^3}) \prod\limits_{i<j}^N (u_i^{\ast}v_j^{\ast}-u_j^{\ast}v_i^{\ast}) ^3{\Phi}_{5} 
\end{equation}
where $(u^{\ast}, v^{\ast})$ is the complex conjugate of the coordinate $(u, v)$, and the $P_{LLL}$ is the lowest Landau level projection operator. 
This can further be rewritten as:
\begin{equation}
\label{EquivalentDerivativePfaffian_Sphere} {\Psi}_{3,5} =  P_{LLL} Pf(\frac{1}{(u_i^{\ast}v_j^{\ast}-u_j^{\ast}v_i^{\ast})^3}) \prod\limits_{i<j}^N |u_iv_j-u_jv_i| ^6{\Phi}_{2} 
\end{equation}
This equivalent form of the wave function suggests ${\Psi}_{3,5}$ can be interpreted as a $f$-wave pairing state of composite femions formed by attaching (due to ${\Phi}_{2}$) two flux quanta to each of the electrons. Similarly, 
${\Psi}_{1,3}$ can be rewritten in the following equivalent form:
\begin{equation}
\label{EquivalentDerivativePfaffian_Sphere} {\Psi}_{1,3} =  P_{LLL} Pf(\frac{1}{u_i^{\ast}v_j^{\ast}-u_j^{\ast}v_i^{\ast}}) \prod\limits_{i<j}^N |u_iv_j-u_jv_i| ^2{\Phi}_{2} 
\end{equation}
This equivalent form of the wave function suggests ${\Psi}_{1,3}$ can be interpreted as a $p$-wave pairing state of composite femions.

Thirdly, as for the Moore-Read Pfaffian state, one can modify the the Pfaffian in Eq.(\ref{DOperator}) to construct the quasihole wave function: 
\begin{equation}
\label{DOperatorQH}{Pf}(\frac{ \prod\limits_{a=1}^n (z_i - {\xi}_a) \prod\limits_{b=n+1}^{2n} (z_j - {\xi}_b) +(i {\leftrightarrow} j)}{ (\frac{\partial}{{\partial}z_i}-\frac{\partial}{{\partial}z_j} )^m})
\end{equation}
for the total number of $2n$ quasiholes located in ${\xi}_a$ and ${\xi}_b$, where $a = 1, 2, \ldots, n$ and $b = n+1, n+2, \ldots, 2n$. The quasiholes carry $-\frac{1}{2(m_L-m)}$ of the electron charge, and obey non-Abelian fractional statistics $\frac{1}{2(m_L-m)}$. The non-Abelian quasiparticles can be formed in a similar way by replacing $z_i$ in Eq.(\ref{DOperatorQH}) with $\frac{\partial}{{\partial}z_i}$.

Finally, since ${\Psi}_{m,m_L}$ is formed by applying coordinate derivatives to the Laughlin wave function, one may
regard it as formed in the quasielectron space of the Laughlin state. It has been shown numerically by Su and the author\cite{YangHierarchy} that the quasielectron space of the Laughlin state provides a rather exact description for the low energy physics from $1/3$ fillling factor all the way to the 2/5 filling factor. Now we have shown that its validity range may go beyond $2/5$ filling factor to $1/2$ filling factor.

\end{document}